# Fluctuation relations for anomalous dynamics

## A V Chechkin[1] and R Klages[2]

[1]Institute for Theoretical Physics NSC KIPT, Akademicheskaya st. 1, Kharkov 61108, Ukraine
[2]Queen Mary University of London, School of Mathematical Sciences, Mile End Road, London E1 4NS, UK

E-mail: achechkin@kipt.kharkov.ua; klages@maths.qmul.ac.uk

**Abstract.** We consider work fluctuation relations (FRs) for generic types of dynamics generating anomalous diffusion: Lévy flights, long-correlated Gaussian processes and time-fractional kinetics. By combining Langevin and kinetic approaches we calculate the probability distributions of mechanical and thermodynamical work in two paradigmatic nonequilibrium situations, respectively: a particle subject to a constant force and a particle in a harmonic potential dragged by a constant force. We check the transient FR for two models exhibiting superdiffusion, where a fluctuation-dissipation relation does not exist, and for two other models displaying subdiffusion, where there is a fluctuation-dissipation relation. In the two former cases the conventional transient FR is not recovered, whereas in the latter two it holds either exactly or in the long-time limit.
**Keywords**: stochastic processes (theory), large deviations in non-equilibrium systems, diffusion, stochastic particle dynamics (theory)

### Introduction

*Fluctuation relations* (FRs) denote large-deviation symmetry properties in probability density functions (PDFs) of nonequilibrium statistical physical observables. One subset of them, *fluctuation theorems*, grew out of generalizations of the second law of thermodynamics to thermostated systems [1,2,3]. Another subset, *work relations*, generalize a thermodynamic equilibrium relation between work and free energy to nonequilibrium situations [4]. These two fundamental classes were generalized by other FRs from which they can partially be derived as special cases [5,6,7]. FRs hold for a great variety of systems thus featuring one of the rare statistical physical principles that is valid very far from equilibrium [8,9]. Many of these relations have been verified in experiments on nanosystems [10,11].

*Anomalous dynamics* refers to processes that do not obey the laws of conventional statistical physics [12,13]. Paradigmatic examples are diffusion processes where the long-time mean square displacement does not grow linearly in time, $\langle x^2(t)\rangle \propto t^\mu$ with $\mu = 1$ for Brownian motion, but either *subdiffusively* with $\mu < 1$ or *superdiffusively* with $\mu > 1$. Such anomalous transport phenomena have recently been observed in a wide variety of complex systems [14]. This raises the question to which extent FRs are valid for anomalous dynamics. Results for generalized Langevin equations [15,16,17,18,19], Lévy flights [20] and continuous-time random walks [21] showed both validity and violations of different FRs.

In this letter we propose to classify FRs for anomalous dynamics by distinguishing between four generic types of anomalous diffusion: We consider a particle exhibiting one-dimensional anomalous diffusion generated by a random force that, firstly, obeys anomalous statistics (Lévy flights) or, secondly, normal statistics but anomalous memory properties (non-Markovian long-correlated Gaussian noise). In the latter case we consider noise that is internal or external depending on the existence of a fluctuation-dissipation theorem. Also, we consider the case described by a time-fractional kinetic equation where anomalous diffusion is stipulated by long power-law asymptotics of the PDF for the random waiting time intervals between instant successive jumps [13].

In all cases, a regular external force given by a potential $U(x, X(t))$ acts on the particle at position $x$, where $X$ is an external control parameter that varies according to a fixed protocol $X(t)$. Following [22], we study our four models in two different nonequilibrium situations: For Class A the particle is driven by a constant external force, for Class B the particle is confined to a moving harmonic potential. We restrict ourselves to *overdamped* motion, where the particle acceleration is negligible. Furthermore, in order to be consistent, we choose the simplest nonequilibrium initial condition $x(t = 0) = x_0 = 0$ for all four cases, since there is no Boltzmann equilibrium for the systems exhibiting Lévy flights and for those driven by an external Gaussian noise.

## Class A. Systems under a constant force

In this Section we consider models driven by a constant external force, $U = -F_0 x$. We are interested in the *mechanical work* PDF $p(W_M, t)$, where the mechanical work $W_M$ is given by $W_M = -\int dx\, \partial U/\partial x = F_0 x$. Note that for Class A systems $W_M$ is identical to the heat. Thus, the PDF $p(W_M, t)$ is simply related to the $x$–PDF $f(x,t)$ by $p(W_M, t) = F_0^{-1} f(W_M/F_0, t)$.

*A1. A system exhibiting Lévy flights.* Our starting point is the Langevin equation for an overdamped Lévy particle moving in a constant field under white Lévy noise,

$$\frac{dx}{dt} = \frac{F_0}{m\gamma} + \xi(t) , \tag{A1}$$

where $F_0$ is a constant force, $m$ the mass, $\gamma$ the friction coefficient, and $\xi(t)$ holds for white Lévy noise. That is, the time integral over $\Delta t$, $L(\Delta t) = \int_{t}^{t+\Delta t} dt'\xi(t')$ is the $\alpha$-stable Lévy process whose PDF $p_\alpha(x, \Delta t)$ has the characteristic function (CF) $\hat{p}_\alpha(k, \Delta t)$ [23],

$$\hat{p}_\alpha(k, \Delta t) = F\{p_\alpha(x, \Delta t)\} \equiv \int_{-\infty}^{\infty} dx\, e^{ikx} p_\alpha(x, \Delta t) = \exp\left[-D_\alpha |k|^\alpha \Delta t\right] , \tag{A2}$$

where $\alpha \in [0, 2]$ is the Lévy index, and $D_\alpha$ has a meaning of the noise intensity. In this paper we restrict ourselves to the case of symmetric Lévy noise, the generalization to asymmetric noise will be given elsewhere. It is well-known that in the absence of an external potential the Lévy particle exhibits superdiffusive motion, in the sense that the fractional moments of the order $\mu$, $0 < \mu < \alpha$, give superdiffusive scaling, $\left\langle |x|^\mu \right\rangle^{2/\mu} \propto t^{2/\alpha}$, that is the "effective second moment" grows faster than $t$ if $\alpha < 2$. The PDF $f(x,t)$ obeys the space-fractional Fokker-Planck equation [13]

$$\frac{\partial f}{\partial t} = -\frac{\partial}{\partial x}\left(\frac{F_0}{m\gamma} f\right) + D_\alpha \frac{\partial^\alpha f}{\partial |x|^\alpha} , \tag{A3}$$

where the Riesz fractional derivative standing on the right hand side is understood via its Fourier transform as $F\{\partial^\alpha f/\partial |x|^\alpha\} = -|k|^\alpha F\{f\} = -|k|^\alpha \hat{f}(k,t)$. Equation (A3) is easily solved in Fourier space, giving for the CF $\hat{p}(k,t)$ of the work PDF, $\hat{p}(k,t) = \hat{f}(kF_0, t)$,

$$\hat{p}(k,t) = \exp\left[\left(i\rho k - \sigma^\alpha |k|^\alpha\right)t\right] , \quad \rho = F_0^2/(m\gamma) , \quad \sigma^\alpha = F_0^\alpha D_\alpha . \tag{A4}$$

Using the CF Eq.(A4), it can easily be seen that $p(W_M, t)$ can be rewritten as

$$p(W_M, t) = \frac{1}{\sigma t^{1/\alpha}} L_\alpha\left(\frac{W_M - \rho t}{\sigma t^{1/\alpha}}\right) , \tag{A5}$$

where $L_\alpha(x)$ is the Lévy stable PDF whose CF is given by $\hat{L}_\alpha(k) = \exp\left(-|k|^\alpha\right)$. It is convenient to introduce *the scaled value of work* $w_M = W_M/(\rho t)$ [20]. We then look at the fraction defining the transient FR

$$g_t(w_M) \equiv \frac{p(W_M,t)}{p(-W_M,t)} = L_\alpha\left(\frac{w_M - 1}{(\sigma/\rho)t^{1/\alpha-1}}\right) \Big/ L_\alpha\left(\frac{-w_M - 1}{(\sigma/\rho)t^{1/\alpha-1}}\right) \quad . \tag{A6}$$

Only when the particle is subjected to a Gaussian noise, $\alpha = 2$, we have a conventional transient FR,

$$g_t(w_M) = \exp(A w_M t) \quad , \tag{A7}$$

where $A = F_0^2/(k_B T m\gamma)$, and we use the Einstein relation, $D_{\alpha=2} = k_B T/(m\gamma)$, with $T$ the temperature of the heat bath and $k_B$ the Boltzmann constant. For arbitrary Lévy noise with $0 < \alpha < 2$ we use the asymptotics of the Lévy stable PDF, $L_\alpha(\xi) \approx C/|\xi|^{1+\alpha}$, $C = \pi^{-1}\sin(\pi\alpha/2)\Gamma(1+\alpha)$ [23], which gives

$$\lim_{w_M \to \pm\infty} g_t(w_M) = 1 \tag{A8}$$

This means that *asymptotically large positive and negative fluctuations of work are equally probable for Lévy flights*. This was established for the first time in the different nonequilibrium situation of a case B system in Ref.[20].

*A2. A system driven by long-correlated internal Gaussian noise.* Let us now consider non-Markovian processes with long-time memory characterized by a memory function exhibiting slow power law decay in time. The starting point is the overdamped Langevin equation (compare with Eq.(A1)),

$$\int_0^t dt'\,\dot{x}(t')K(t-t') = \frac{F_0}{m\gamma} + \xi(t) \quad , \tag{A9}$$

where the dot above $x$ denotes the time derivative. The autocorrelation function of the Gaussian noise is connected with the friction kernel by the fluctuation-dissipation relation of the second kind [24] $\langle \xi(t)\xi(t')\rangle = (k_B T/m\gamma)K(t-t')$, which implies that we treat $\xi(t)$ as an *internal* noise. To model long – time memory, a natural choice for the friction kernel is $K(t) = \tau_\beta^{\beta-1} t^{-\beta}/\Gamma(1-\beta), t \geq 0, 0 < \beta < 1$. Here, by including the factor $1/\Gamma(1-\beta)$, we may use the limit $t^{-\beta}/\Gamma(1-\beta) \to 2\delta(t), \beta \to 1^-$, to obtain $\langle \xi(t)\xi(t')\rangle = 2k_B T\delta(t-t')/(m\gamma)$, thus recovering the case of overdamped (ordinary) Brownian motion. Equation (A.9) is easily solved in Laplace space, $\tilde{x}(s) = \int_0^\infty dt\, x(t)e^{-st}$, giving after the inverse Laplace transformation

$$x(t) = \frac{F_0 \tau_\beta^{1-\beta}}{m\gamma}\frac{t^\beta}{\Gamma(1+\beta)} + \int_0^t dt'\,\xi(t')H(t-t') \quad , \tag{A10}$$

where $H(t) = (t/\tau)^{\beta-1}/\Gamma(\beta)$. The $x$-PDF is Gaussian, and thus the work PDF is also Gaussian, with mean and variance given by

$$\langle W_M\rangle = F_0 \langle x(t)\rangle = \frac{F_0^2 \tau_\beta^{1-\beta}}{m\gamma}\frac{t^\beta}{\Gamma(1+\beta)} \quad , \quad \sigma_W^2 = F_0^2\left\langle (x(t)-\langle x(t)\rangle)^2\right\rangle = \frac{2\tau_\beta^{1-\beta}F_0^2}{\Gamma(1+\beta)}\frac{k_B T}{m\gamma}t^\beta \quad . \tag{A11}$$

From the second formula of (A.11) it follows that the particle exhibits *subdiffusion*. Thus, from Eq.(A11) we conclude that the *subdiffusion dynamics caused by long-correlated Gaussian noise in presence of the fluctuation-dissipation theorem of the second kind leads to a conventional transient FR*,

$$p(W_M,t)/p(-W_M,t) = \exp\{W_M/(k_B T)\} \quad . \tag{A12}$$

*A3. A system driven by long-correlated external Gaussian noise.* The starting point is again the Langevin equation Eq.(A1), however, we now assume that $\xi(t)$ is a stationary Gaussian process with zero mean, $\langle \xi(t)\rangle = 0$, and autocorrelation function,

$$\langle \xi(t)\xi(t')\rangle = \frac{C_\beta}{\Gamma(1-\beta)\gamma^2}|t-t'|^{-\beta} \quad , \quad 0<\beta<1 \quad , \tag{A13}$$

where $C_\beta$ is a constant. Here the noise $\xi(t)$ is treated as an *external* noise, since contrary to the case A2 the fluctuation-dissipation theorem of the second kind is not valid in this system. As $\beta \to 1$ and $C_1 = \gamma k_B T/m$, we obtain $\langle \xi(t)\xi(t')\rangle = 2k_B T \delta(t-t')/(m\gamma)$, and Eq.(A1) together with Eq.(A.13) boils down to the Langevin description of an overdamped Brownian particle.

The work PDF is easily constructed as a Gaussian function with mean $\langle W_M(t)\rangle = F_0^2 t/(m\gamma)$ and variance $\sigma_W^2 = 2C_\beta t^{2-\beta} F_0^2/\left(\gamma^2 \Gamma(3-\beta)\right)$. We note that the mean square displacement grows as $t^{2-\beta}$, that is, the system exhibits *superdiffusion*, in contrast to the internal noise case. Here it is convenient to introduce the *mean production of heat per unit time*, $\mu_W = \langle W_M\rangle/t = F_0^2/(m\gamma)$, and the *scaled value of work*, $w_M = W_M/\langle W_M\rangle$. The transient FR for the heat then takes the form

$$g_t(w) = p(W_M,t)/p(-W_M,t) = \exp\left(A(\beta)w_M t^\beta\right) \quad , \tag{A14}$$

where $A(\beta) = \Gamma(3-\beta)\gamma\mu_W/(mC_\beta)$. Equation (A14) tells us that the *superdiffusion dynamics caused by external long-correlated Gaussian noise leads to a "non-conventional" transient FR of stretched exponential type*. To our knowledge, this is the first time that a FR has been derived that still reproduces the exponential form of conventional FRs by containing an explicit time-dependence with a fractional power of time. As $\beta \to 1$ and $C_1 = \gamma k_B T/m$ we arrive at the conventional transient FR for a Brownian particle. Similar results have been obtained for a random walk model with memory-dependent transition rates by applying functional integration techniques [25].

*A4. A system described by a time-fractional kinetic equation.* The starting point is a set of coupled Langevin equations for the motion of a particle [26,27]

$$\frac{dx(u)}{du} = \frac{F_0}{m\gamma} + \xi(u) \quad , \quad \frac{dt(u)}{du} = \tau(u) \quad , \tag{A15}$$

where the random walk $x(t)$ is parametrized by the variable $u$. The random process $\xi(u)$ is a white Gaussian noise, $\langle \xi(u)\rangle = 0$, $\langle \xi(u)\xi(u')\rangle = 2k_B T\delta(u-u')/(m\gamma)$, and $\tau(u)$ is a white Lévy stable noise, which takes positive values only and obeys a totally skewed alpha-stable Lévy distribution with $0<\alpha<1$. It was demonstrated [26,27] that such a *subordinated* Langevin description is equivalent to the time-fractional Fokker-Planck equation

$$\frac{\partial f}{\partial t} = D_t^{1-\alpha}\left[-\frac{\partial}{\partial x}\frac{F_0}{m\gamma_\alpha} + K_\alpha\frac{\partial^2}{\partial x^2}\right]f(x,t) \quad , \quad f(x,t=0) = \delta(x) \quad , \tag{A16}$$

which is used to model a variety of subdiffusion phenomena, see, e.g., [13] for detailed discussions. In this equation $\gamma_\alpha$ and $K_\alpha$ are generalized friction and diffusion constants, respectively, obeying the (generalized) Einstein relation $K_\alpha = k_B T/(m\gamma_\alpha)$, and $D_t^{1-\alpha}$ is the Riemann – Liouville fractional derivative on the right semi-axis, which, for a "sufficiently well-behaved" function $\phi(t)$ is defined as $D_t^\mu\phi = \Gamma^{-1}(1-\mu)(d/dt)\int_0^t d\tau(t-\tau)^{-\mu}\phi(\tau)$, $0 \leq \mu < 1$, with Laplace transform $s^\mu\tilde{\phi}(s)$. From Eq.(A16) the equations for the first and the second moments can easily be obtained and then solved by the Laplace transformation. The mean square displacement in the absence of any external force is given by $\langle x^2(t)\rangle_0 = 2K_\alpha t^\alpha/\Gamma(1+\beta)$ demonstrating *subdiffusive* behavior. We note also that the (second) Einstein relation is recovered, $\langle x(t) \rangle_{F_0} = F_0 \langle x^2(t) \rangle_0 / (2k_B T)$, which connects the first moment in presence of a constant force $F_0$ with the second moment in the absence of this force [13]. Both Einstein relations are fluctuation-dissipation relations of the first kind for this system [24].

Applying the Laplace transform to Eq.(A.16), and solving the equation in the Laplace space separately for $x > 0$ and $x < 0$, we get, with $\tilde{f}(x,s) \to 0$ at $x \to \pm\infty$,

$$\tilde{f}(x,s) = \frac{s^{\alpha-1}}{\sqrt{V_0^2 + 4K_\alpha s^\alpha}} \exp\left( \frac{V_0 x}{2K_\alpha} - |x| \frac{\sqrt{V_0^2 + 4K_\alpha s^\alpha}}{2K_\alpha} \right) , \quad (A17)$$

where $V_0 = F_0 / m\gamma$. Note that at $\alpha = 1$ Eq.(A17) gives the Laplace transform of the Gaussian distribution. In the general case of $0 < \alpha < 1$ we have for the ratio of the Laplace transforms for the work PDFs,

$$\frac{\tilde{p}(W_M, s)}{\tilde{p}(-W_M, s)} = \frac{\tilde{f}(W_M / F_0, s)}{\tilde{f}(-W_M / F_0, s)} = \exp\left( \frac{W_M}{k_B T} \right) . \quad (A18)$$

Transferring $\tilde{p}(-W_M, s)$ from the left hand side to the right hand side of Eq.(A18) and then making an inverse Laplace transformation, we arrive at the FR in the time domain. Thus, we conclude that, similar to the case A2 with long-correlated internal Gaussian noise, *subdiffusive dynamics modeled by a time-fractional Fokker-Planck equation obeying a fluctuation-dissipation relation leads to a conventional transient FR*.

### Class B. Systems coupled to a harmonic oscillator

In this Section we consider a particle confined by a harmonic potential that is dragged by a constant velocity, $U = (\kappa/2)(x - X(t))^2$, where $X(t) = v_* t$, $v_* = $ const. We are interested in the PDF of *thermodynamical work* $W_T$ given by

$$W_T(t) = \int dX \, \partial U / \partial X = \int_0^t dt' (dX(t')/dt') \partial U / \partial X = -\kappa v_* \int_0^t dt' (x - v_* t') . \quad (B1)$$

***B1. A system exhibiting Lévy flights.*** The starting point is the coupled Langevin equations written in the comoving coordinate frame, $y = x - v_* t$,

$$\frac{dy}{dt} = -v_* - \frac{1}{\tau_*} y + \xi(t) \quad , \quad \frac{dW_T}{dt} = -\kappa v_* y(t) , \quad (B2)$$

where $\xi(t)$ is a white Lévy noise as in Section A1, $\tau_* = m\gamma / \kappa$ is relaxation time. For this case the CF of the work PDF was calculated in [20] by using a functional integration technique. We propose here a different approach based on the generalized space-fractional kinetic equation for the joint PDF $\phi(y, W_T, t)$ (or $\phi(x, W_T, t)$). The kinetic equation for this PDF can be constructed almost immediately from noticing that, with the proper change of variables, Eqs.(B2) define the Langevin equations for the underdamped Lévy particle, for which $y$ and $W$ have the meaning of velocity and coordinate, respectively. The corresponding kinetic equation is known in the theory of Lévy flights as a velocity-fractional Klein-Kramers equation [28]. Thus, we have

$$\frac{\partial}{\partial t} \phi(y, W_T, t) - v_* \frac{\partial \phi}{\partial y} = \frac{1}{\tau_*} \frac{\partial}{\partial y}(y\phi) + D_\alpha \frac{\partial^\alpha \phi}{\partial |y|^\alpha} + \kappa v_* y \frac{\partial \phi}{\partial W_T} . \quad (B3)$$

Equation (B3) is subject to the initial condition $\phi(y, W_T, t=0) = \delta(y)\delta(W_T)$. We note that at $\alpha = 2$ Eq.(B3) corresponds to the equation for the PDF $\phi(y, W_T, t)$ of a driven Brownian particle [29].

To solve Eq.(B3) we make a double Fourier transformation, $\hat{\phi}(k,q,t) = \int_{-\infty}^{\infty} dy \int_{-\infty}^{\infty} dW_T \exp(iky + iqW_T)\phi(y,W_T,t)$, and solve the equation for the CF $\hat{\phi}(k,q,t)$ by the method of characteristics. We present here a simpler CF of the work PDF,

$$\ln \hat{p}(q,t) = \ln \hat{\phi}(k=0,q,t) = iqA - |q|^\alpha B(\alpha) \ , \qquad (B4)$$

where

$$A = v_*^2 \tau_*^2 \kappa \left( \frac{t}{\tau_*} - 1 + e^{-t/\tau_*} \right) \ , \quad B_\alpha = D_\alpha v_*^\alpha (m\gamma)^\alpha \int_0^t dt' \left( 1 - e^{-(t-t')/\tau_*} \right)^\alpha \ . \qquad (B5)$$

The result given by Eqs.(B4) and (B5) is identical to that reported in Ref.[20]. As a consequence, for the work PDF of the Lévy flights we have the same relation as that derived previously for the heat PDF in case of a constant force, Eq.(A10), which means that *asymptotically large positive and negative fluctuations of thermodynamical work are equally probable for Lévy flights.*

***B2. A system driven by long-correlated internal Gaussian noise.*** The starting Langevin equation in the comoving coordinate frame has the form

$$-\frac{y}{\tau_*} - \int_0^t dt' \dot{y}(t') K(t-t') - v_* \int_0^t dt' K(t') + \xi(t) = 0 \ , \qquad (B6)$$

where $K(t)$ is related to $\langle \xi(t)\xi(t') \rangle$ via the fluctuation-dissipation relation of the second kind, see Section A2. Similar problems have been studied in Refs.[17,18]: In Ref.[17] an underdamped oscillator driven by internal fractional Gaussian noise was considered, Ref.[18] analyzes an overdamped oscillator but with equilibrium initial condition.

Equation (B6) is easily solved in Laplace space. Taking into account the Laplace transformation, $\int_0^\infty dt\, e^{-st} t^{b-1} E_{a,b}(-ct^a) = s^{a-b}/(s^a + c)$, where $E_{a,b}(z) = \sum_{k=0}^\infty z^k / \Gamma(ak+b)$ is a Mittag-Leffler function in two parameters, whose exhaustive list of properties can be found, for example, in [30], Eq.(B6) gives (recall that $y(0) = 0$)

$$W_T(t) = -\kappa v_* \int_0^t dt' H_1(t-t') \xi(t') + \kappa v_*^2 H_2(t) \ , \qquad (B7)$$

where

$$H_1(t) = \tau_* \left[ 1 - E_{\beta,1}\left(-ct^\beta\right) \right] \ , \quad H_2(t) = t^2 E_{\beta,3}\left(-ct^\beta\right) \ , \qquad (B8)$$

and $c = \tau_\beta^{1-\beta}/\tau_*$. Using the relation $\dot{H}_2(t) = \int_0^t dt' H_1(t') K(t-t')$, which can be easily checked in Laplace space, we get the work PDF, which is Gaussian with mean and variance given, respectively, by

$$\langle W_T \rangle = \kappa v_*^2 H_2(t) \ , \quad \sigma_{W_T}^2 = 2\kappa^2 v_*^2 \frac{k_B T}{m\gamma} \int_0^t d\tau H_1(\tau) \dot{H}_2(\tau) \ . \qquad (B9)$$

Using first the formulas of derivative and integral of the Mittag–Leffler function, see Ref.[30] Eqs. (1.83) and (1.99), respectively, and second the asymptotics $E_{a,b}(z) \approx -z^{-1}/\Gamma(b-a)$ (ibid, Eq.(1.143)), we get *asymptotically* at $t \to \infty$ the conventional FR, $p(W_T,t)/p(-W_T,t) = \exp\{W_T/(k_B T)\}$. We note that in contrast to the case with a constant force, Section A2, the conventional fluctuation relation holds here in the asymptotic limit of long times only.

***B3. A system driven by long-correlated external Gaussian noise.*** The starting point is again the Langevin equation in the comoving coordinate frame, Eqs.(B2), where we assume that $\xi(t)$ is a stationary Gaussian process with zero mean, $\langle \xi(t) \rangle = 0$, and a pair correlation function given by Eq.(A13). Solving the first equation of Eqs.(B2) with the initial condition $y(0) = 0$, we get

$$y(t) = -v_* \tau_* \left(1 - e^{-t/\tau_*}\right) + \int_0^t dt' \xi(t') \exp\left(-\frac{t-t'}{\tau_*}\right) \quad . \tag{B10}$$

Using the second equation from Eqs.(B2), we get an expression for the work $W_T$ and then construct the work PDF as the Gaussian function with the mean $\langle W_T \rangle$ given by the term $A$ in Eq.(B5) and the variance

$$\sigma_{W_T}^2 = \frac{m^2 v_*^2 C_\beta t^{2-\beta}}{\Gamma(3-\beta)} \left\{ 2 - e^{-t/\tau_*}\left(2 - e^{-t/\tau_*}\right) M\left(2-\beta, 3-\beta; \frac{t}{\tau_*}\right) - e^{-t/\tau_*} M\left(1, 3-\beta, t/\tau_*\right) \right\} , \tag{B11}$$

where $M(a,b,z)$ is a Kummer's function. At $\beta = 1$ and $C_1 = \gamma k_B T/m$ Eq.(B11) yields the result for the Brownian motion with non-equilibrium initial condition $x(0) = 0$. After the relaxation stage, $t \gg \tau_*$, we have for the mean and variance of the work, respectively,

$$\langle W_T \rangle = m v_*^2 \gamma t , \quad \sigma_{W_T}^2 = \frac{2}{\Gamma(3-\beta)} m^2 v_*^2 C_\beta t^{2-\beta} \quad . \tag{B12}$$

Similar to Section A, we introduce a mean production of work $\nu$ per unit time at $t \gg \tau_*$, $\langle W_T \rangle / t \equiv \nu = \gamma m v_*^2 = const$, as well as a scaled value of work $w$, $w_T = W_T / \langle W_T \rangle$, $W_T = \nu w_T t$. With that we get the transient FR in the form

$$p(W_T, t) / p(-W_T, t) = \exp\left(B(\beta) w_T t^\beta\right) \quad , \tag{B13}$$

where $B(\beta) = \Gamma(3-\beta) \gamma^2 v_*^2 / C_\beta$. This agrees with the FR for the heat, Eq.(A14). The conventional FR is recovered in the limit of $\beta = 1$.

**B4. A system described by a time-fractional kinetic equation.** Similar to Section A4, the starting point is the coupled Langevin equations written in a comoving frame as

$$\frac{dy(u)}{du} = -v_* - \frac{\kappa}{m\gamma} y(u) + \xi(u) \quad , \quad \frac{dt(u)}{du} = \tau(u) \quad , \quad \frac{dW_T}{dt} = -\kappa v_* y(t) \quad , \tag{B14}$$

where we have added the equation for the work $W_T$. Now, we are able to construct a generalized fractional kinetic equation governing the joint PDF for the work and coordinate. Indeed, introducing $w_T = -W_T / (\kappa v_*) = W_T / (\kappa V_*)$, $V_* = -v_*$, we observe that the system (B14) is equivalent to that considered by Friedrich and co-workers in connection with fractional kinetic equations including inertial effects [31], if $w$ and $y$ are regarded, respectively, as coordinate and velocity of the inertial particle. This set of Langevin equations is equivalent to the fractional Kramers – Fokker – Planck equation proposed in Ref.[32]. In our notation

$$\left(\frac{\partial}{\partial t} + y\frac{\partial}{\partial w_T} + V_*\frac{\partial}{\partial y}\right)\phi(w_T, y, t) = \left(\frac{\kappa}{m\gamma_\alpha}\frac{\partial}{\partial y} y + K_\alpha \frac{\partial^2}{\partial y^2}\right) D_t^{1-\alpha} \phi(w_T, y, t) \quad , \tag{B15}$$

where $\gamma_\alpha$ and $K_\alpha$ are generalized friction and diffusion constants, respectively, as in Section A4, and $D_t^{1-\alpha}$ is a fractional substantial derivative defined as

$$D_t^{1-\alpha} \phi(w_T, y, t) =$$

$$\left(\frac{\partial}{\partial t} + y\frac{\partial}{\partial w_T} + V_*\frac{\partial}{\partial y}\right) \frac{1}{\Gamma(\alpha)} \int_0^t \frac{dt'}{(t-t')^{1-\alpha}} \exp\left[-(t-t')\left(y\frac{\partial}{\partial w_T} + V_*\frac{\partial}{\partial y}\right)\right] \phi(w_T, y, t'). \tag{B16}$$

The solution can be obtained by following the method developed in [33]. After getting the solution of this equation it is possible to check the FR for the work PDF $\phi(W_T, t)$, as will be discussed in detail in a long paper.

**Conclusions**

We have shown that for two superdiffusive systems without fluctuation-dissipation relation, one subject to white Lévy stable noise and the other one to long-correlated external Gaussian noise, the conventional transient FR does not hold. Namely, by applying two methods, a Langevin approach and one based on a space – fractional kinetic equation, we have found that for stochastic systems driven by Lévy noise the asymptotically large positive and negative fluctuations of work are equally probable, which generalizes previous studies in Ref.[20] of the thermodynamical work fluctuation theorem for Lévy flights. For the systems driven by long-correlated external Gaussian noise we found a new, unconventional FR characterized by stretched exponential type behavior in time. On the other hand, for two subdiffusive systems with fluctuation-dissipation relation, one subject to long-correlated internal Gaussian noise and the other one modeled by a time-fractional kinetic equation, the conventional transient FR is recovered. To our knowledge, this is the first time that the transient FR is verified for time-fractional kinetics. Our studies of these four generic types of anomalous dynamics suggest an intimate connection between fluctuation-dissipation relations and FRs in case of anomalous diffusion.

We expect our results to have important applications to experiments: Recently it has been shown that migrating biological cells exhibit anomalous dynamics similar to that under the influence of correlated Gaussian noise [34]. This suggests to check whether cells migrating under chemical concentration gradients obey anomalous FRs. A second type of experiments would be to drag a particle through a highly viscous gel instead through water [10], or to measure the fluctuations of a driven pendulum in gel [35]. Thirdly, one may check for anomalous FRs for granular gases exhibiting subdiffusion dynamics [36]. On the theoretical side, our approach paves the way to systematically check the remaining variety of conventional FRs [5,6,7] for anomalous generalizations.

We thank H.Touchette for his careful reading of the manuscript and both him and R.J. Harris for helpful discussions. Financial support from EPSRC under grant no EP/E00492X/1 is gratefully acknowledged.We thank H.Touchette for his careful reading of the manuscript and both him and R.J. Harris for helpful discussions. Financial support from EPSRC under grant no EP/E00492X/1 is gratefully acknowledged.